\documentclass[aps,prx,floatfix,twocolumn,superscriptaddress]{revtex4-1}
\usepackage{amsmath,amssymb}
\usepackage{graphicx}
\usepackage{epsfig}
\usepackage{psfrag}
\usepackage[usenames]{color}
\usepackage{xcolor}
\usepackage{hyperref}
\usepackage{soul,color}
\hypersetup{colorlinks=true,citecolor={blue},linkcolor={blue},urlcolor={blue}}

\begin{document}

\preprint{APS/123-QED}

\title{Coulomb drag of excitons in Bose-Fermi mixtures}

\author{M.~V.~Boev}
 \affiliation{A.V.~Rzhanov Institute of Semiconductor Physics, Siberian Branch of Russian Academy of Sciences, Novosibirsk 630090, Russia}

\author{V.~M.~Kovalev}
\affiliation{A.V.~Rzhanov Institute of Semiconductor Physics, Siberian Branch of Russian Academy of Sciences, Novosibirsk 630090, Russia}
\affiliation{Department of Applied and Theoretical Physics, Novosibirsk State Technical University, Novosibirsk 630073, Russia}

\author{I.~G.~Savenko}
\affiliation{Center for Theoretical Physics of Complex Systems, Institute for Basic Science (IBS), Daejeon 34126, Korea}
\affiliation{A.V.~Rzhanov Institute of Semiconductor Physics, Siberian Branch of Russian Academy of Sciences, Novosibirsk 630090, Russia}

\date{\today}

\begin{abstract}
We develop a microscopic theory of the Coulomb drag effect in a hybrid system consisting of spatially separated two-dimensional quantum gases of degenerate electrons and dipolar excitons. 
We consider both the normal-phase and condensate regimes of the exciton subsystem and investigate the cross-mobility of the system being the kinetic coefficient, which couples the static electric field applied to the electron layer with the particle density current (flux) in the exciton subsystem. 
We study the temperature dependence of the cross-mobility and its dependence on the interlayer separation. 
We show that exciton-exciton interaction plays a dramatic role. 
If the exciton gas is in the normal phase, then the screening of interlayer interaction by the exciton subsystem results in an exponential damping of the cross-mobility with the decrease of temperature, while at low temperatures, the interactions result in a robust bosonic transport due to the emergence of the Bogoliubov quasiparticles.
\end{abstract}

\maketitle


\section{\label{sec:level1}Introduction}
The nature of the Coulomb drag effect (CDE) is in the interaction between components of a complex system, which results in a particle current in one of the subsystems called passive due to the presence of a particle current in the other subsystem called active. This phenomenon has been broadly studied in various structures~\cite{RefNarozhnyRMP}, such as two-dimensional electron gases (2DEGs) located in spatially separated quantum wells, monoatomic layers of graphene, and hybrid Bose--Fermi systems (BFSs).

The latter have recently attracted considerable interest. They represent a research platform for both the fundamental effects resulting from many-body interaction phenomena and  applications~\cite{RefImamogluPRB2016}. For instance, in cold atomic gases, using BFSs allows a change of the properties of the Feshbach resonance~\cite{RefFesh1, RefFesh2, RefFesh3, RefFesh4}, thus tuning the inter-atomic interaction~\cite{RefPhase1, RefPhase2}.
In the solid-state physics, BFSs consist of direct or dipolar exciton or exciton-polariton gases in normal or condensed phase~\cite{RefShelykhPRL2010, RefButov, RefTimofeev}, which interact with electrons and holes residing in the same or separate layers. 
Such interactions result in various curious phenomena, including the solid-state Fano resonance~\cite{OurFano}, formation of the exciton supersolid phase~\cite{RefShelykhPRL1051404022010, RefMatuszewskiPRL1080604012012}, and opening new mechanisms of scattering~\cite{OurCapture}.

A typical BFS is a semiconductor heterostructure, hosting a 2DEG which resides in a layer of a metal or n-doped semiconductor and an indirect exciton gas occupying two parallel quantum wells.
The CDE has been studied in such systems~\cite{Lozovik_1, Lozovik_2, Lozovik_3} to some extent. There electrons represent an active layer and the excitons are the passive one. The theory developed in these works is based on the quasi-classical Boltzmann equations approach, which is not applicable for treating of vertex corrections resulting from multiple scattering processess.

In this paper, we develop a quantum microscopic theory of the CDE in hybrid 2DEG -- dipolar exciton gas system. For that we use the Green's functions approach~\cite{Kamenev_Oreg}, which allows us to consider both the ballistic and diffusive regimes of particle motion in normal and condensed phases of the bosonic subsystem with account of the screening effects. 

In Refs.~\onlinecite{Lozovik_1, Lozovik_2}, the interaction potential between excitons and electrons was considered screened by the electrons only, thus disregarding the contribution of the bosonic subsystem into the dielectric permittivity. 
However, despite the neutrality of excitons, their screening might play an important role both in normal and bose-condesed phases of exciton gas~\cite{K_Ch_1}. 
We will show that in the normal phase the exciton contribution to dielectric permittivity essentially modifies the temperature dependence of the cross-mobility, which is a retarded correlation function ``exciton flux -- electric current''.
On the other hand, at low temperatures, the inter-exciton interaction causes the response function of condensed excitons to be very peaked at the eigenfrequency of the elementary excitations from the condensate.


\section{\label{sec:level2}Cross-mobility of an exciton-electron system}
\begin{figure}[t!]
\includegraphics[width=0.99\linewidth]{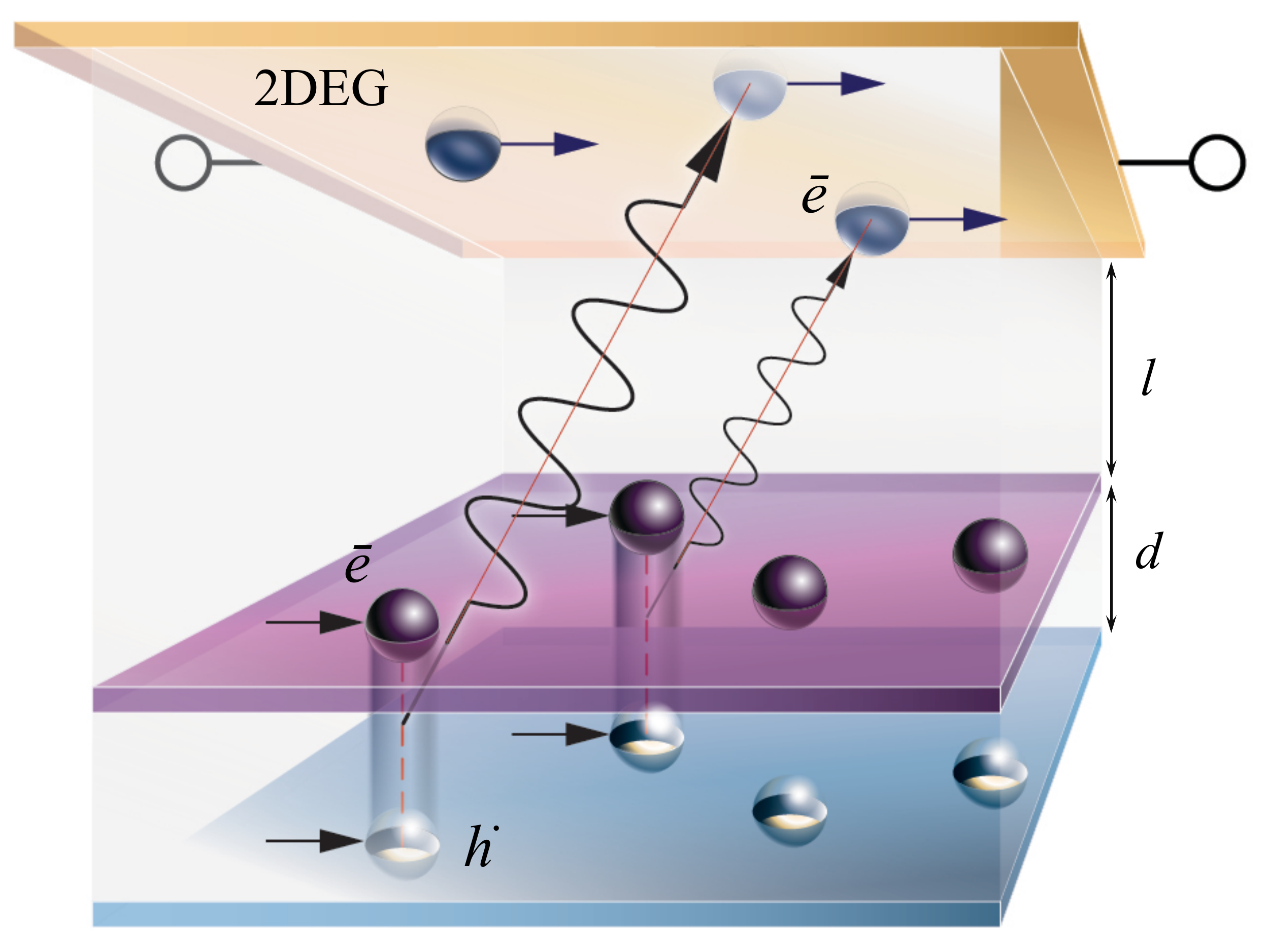}
\caption{System schematic: Hybrid 2DEG--indirect exciton gas system. The electron layer is separated from exciton one by the distance $l$. The distance between electrons and holes in exciton layer is equal to $d$. An electric field $\mathbf{E}$ makes the electrons move. In turn, the electrons drag the excitons due to the Coulomb interaction.} 
\label{Fig1}
\end{figure}

We consider a hybrid 2DEG--indirect exciton gas system (Fig.~\ref{Fig1}). 
%
%
%
The electron-exciton interaction results in the exciton flux, provided that an electric field is applied  to the electronic layer. 
Within the linear response theory, the exciton flux $\textbf{J}=n_{ex}\mu_D\textbf{E}$ is proportional to the external (bias) electric field $\textbf{E}$, the density of exciton gas $n_{ex}$, and the cross-mobility $\mu_D$, which we will concentrate on.
The static cross-mobility reads
\begin{gather}\label{eq.I.1}
\mu_D=\lim\limits_{\Omega\rightarrow 0}\lim\limits_{\textbf{Q}\rightarrow 0}\mu_D(\textbf{Q},\Omega),
\end{gather}
where  the Fourier transform of the cross-mobility, $\mu_D(\textbf{Q},\Omega)$, is defined by the Kubo formula~\cite{Mahan}:
\begin{gather}\label{eq.I.2}
n_{ex}\mu_D(\textbf{Q},\Omega)= \frac{-e}{\Omega} \int d(t-t')d(\textbf{R}-\textbf{r})\Theta(t-t')
\\\nonumber
\times e^{-i\textbf{Q}(\textbf{R}-\textbf{r})+i\Omega(t-t')}
\left\langle\left[\hat{\textbf{J}}(\textbf{R},t), \hat{\textbf{j}}(\textbf{r},t')\right]\right\rangle.
\end{gather}
Here $-e\hat{\textbf{j}}(\textbf{r},t')$ is the operator of the electric current density in the 2DEG layer, $\hat{\textbf{J}}(\textbf{R},t)$ is the operator of the exciton flux, $\textbf{R}$ and $\textbf{r}$ are the coordinates of exciton center of mass and electron respectively and $\Theta(t)$ is the Heaviside step function. 
We will consider the indirect excitons as rigid dipoles, meaning that we will account for the motion of their center of mass only and disregard the excitation of their internal degrees of freedom. 
Following a standard calculation based on the Matsubara technique, let us consider a correlation function
\begin{gather}\label{eq.I.3}
\Pi(\textbf{R}-\textbf{r},\tau-\tau')=-\langle T_\tau S(\beta)\hat{\textbf{J}}(\textbf{R},\tau) \hat{\textbf{j}}(\textbf{r},\tau')\rangle,
\end{gather}
where $\beta=1/T$ is the inverse temperature and $T_\tau$ is the operator of imaginary-time ordering (we put $k_B=\hbar=1$ in what follows). In Eq.~(\ref{eq.I.3}) we omitted the denominator $\langle  S(\beta)\rangle$, implying that only the connected diagrams should be accounted for. Expanding the S-matrix $S(\beta)=T_\tau exp[-\int_0^\beta d\tau \, H'(\tau)]$ into series over the operator of the exciton-electron interaction 
\begin{equation}\label{eq.I.4}
\hat{H}'(\tau)=\int d\textbf{r}d\textbf{R} \, V(\textbf{R}-\textbf{r})\hat{n}(\textbf{R},\tau) \hat{\rho} (\textbf{r},\tau),
\end{equation}
where $V(\textbf{R}-\textbf{r})$ is the electron-exciton interaction energy and $\hat{n}(\textbf{R},\tau)$ and $\hat{\rho} (\textbf{r},\tau)$ are operators of exciton and electron particle densities, we come up with the second-order term in the expansion of the correlation function~(\ref{eq.I.3}):
\begin{eqnarray}\label{eq.I.5}
&&\Pi^{(2)}(\textbf{R}-\textbf{r},\tau-\tau')=\\
\nonumber
&&~~~=-\frac{1}{2} \int_0^\beta d\tau_1 d\tau_2\langle T_\tau\hat{H}'(\tau_1)\hat{H}'(\tau_2)\hat{\textbf{J}} (\textbf{R},\tau) \hat{\textbf{j}}(\textbf{r},\tau')\rangle.
\end{eqnarray}
This term is the lowest-order non-zero term since the first-order contribution $\Pi^{(1)}$ vanishes in the static limit~(\ref{eq.I.1}).

Further we note that the electronic and excitonic degrees of freedom are decoupled  in each term of the perturbation expansion of~(\ref{eq.I.3}). 
It means that the thermal average in Eq.~(\ref{eq.I.5}) should be performed independently for both the 2DEG and exciton gas. Thus it is natural to introduce nonlinear response functions in the following way:
\begin{eqnarray}\label{eq.I.5b}
&&\mathbf{\Delta}_{ex}=-\langle T_\tau\hat{\textbf{J}} (\textbf{R},\tau)\hat{n}(\textbf{R}_1,\tau_1) \hat{n}(\textbf{R}_2,\tau_2)\rangle,
\\\nonumber
&&\mathbf{\Delta}_{e}=-\langle T_\tau\hat{\textbf{j}}(\textbf{r},\tau')\hat{\rho} (\textbf{r}_1,\tau_1) \hat{\rho} (\textbf{r}_2,\tau_2)\rangle.
\end{eqnarray}
Below we will have to deal with the Fourier transforms of these functions. 
Since the system under study possesses the temporal and spatial (in-plane) translational invariance,
the Fourier transforms of Eqs.~\eqref{eq.I.5b} represent the functions of four arguments: $\mathbf{\Delta}=\mathbf{\Delta} (\mathbf{q}_1,\mathbf{q}_2;i\omega_n,i\omega_m)$, where $i\omega_n=2\pi in/\beta$ is an even Matsubara frequency. Furthermore the uniformity of the external electric field additionally reduces the number of momentum arguments, thus $\mathbf{q}_1=\mathbf{q}_2\equiv \mathbf{q}$.

Applying the Fourier transform to Eq.~(\ref{eq.I.5}), we find
\begin{equation}\label{eq.I.6}
\begin{split}
\Pi^{(2)}(Q=0,i\Omega_n)&=
\\
-\frac{1}{2}\sum\limits_\textbf{q}\frac{1}{\beta} \sum_{i\omega_m}\, &V(\textbf{q},i\omega_m) V(\textbf{q},i\omega_m+i\Omega_n)
\\
\times\, \mathbf{\Delta}_{ex}(\textbf{q};i\Omega_n+&i\omega_m,i\omega_m) \mathbf{\Delta}_{e}(\textbf{q};i\omega_m,i\Omega_n+i\omega_m),
\end{split}
\end{equation}
where $V(\textbf{q},i\omega_m)$ is the screened interlayer exciton-electron interaction. 
First, we can perform the summation over the boson frequencies $i\omega_n$. 
A common approach is to switch from the sum to a contour integral $\beta^{-1} \sum_{i\omega_n}\rightarrow  (2\pi i)^{-1}\oint dz n_B(z)$. 
In present case, there are two branch cuts: with $\textrm{Im}(z)=0$ and $\textrm{Im}(z)=-i\Omega_n$ in the complex plane. 
In previous works \cite{Kamenev_Oreg,Flensberg} it has been shown, that only the region of the complex plane inclosed between these branch cuts contributes to the integral in the limit $\Omega\rightarrow 0$.
Second, we note, that the Fourier transform of the cross-mobility~\eqref{eq.I.2} is a retarded function. Therefore the analytic continuation $i\Omega_n\rightarrow \Omega+i\delta$ is necessary to combine Eq.~\eqref{eq.I.6} with Eq.~(\ref{eq.I.2}). 
As a result we find the general expression for the static cross-mobility:
\begin{eqnarray}\label{eq.I.9}
\mu_D&=&\frac{e}{2n_{ex}}\, \int\limits_{-\infty}^{\infty} \frac{d\omega}{2\pi}\frac{\partial n_B(\omega)}{\partial\omega}\times
\\\nonumber
&&\times\sum\limits_\textbf{q}\, \mathbf{\Delta}_{ex}(\textbf{q};\omega^+,\omega^-) \left|\frac{V_q}{\epsilon^R(\textbf{q},\omega)}\right|^2 \mathbf{\Delta}_{e}(\textbf{q};\omega^-,\omega^+),
\end{eqnarray}
where $\omega^{\pm}=\omega\pm i\delta$, $n_B(\omega)=1/(e^{\omega/T}-1)$ is the Bose-Einstein distribution, and the bare electron-exciton interaction reads
\begin{gather}\label{eq.I.10}
V_q=\frac{2\pi e^2d}{\epsilon}e^{-ql}.
\end{gather}
Here $\epsilon$ is the permittivity of the medium and $\epsilon^R(\textbf{q},\omega)$ is the dielectric function, describing the screening.

The presence of impurities in the sample requires averaging~(\ref{eq.I.9}) over their positions. 
Let us assume that the electron-impurity and exciton-impurity scattering events occur independently, so that $\overline{\mathbf{\Delta}_{ex}\mathbf{\Delta}_{e}} \approx \overline{\mathbf{\Delta}}_{ex}\overline{\mathbf{\Delta}}_{e}$. 
In addition, we suppose that the random impurity field $u_\alpha(\textbf{r})$ ($\alpha=e,ex$) satisfies the following white-noise correlations:
\begin{gather}\label{eq.I.12}
\langle u_\alpha\rangle=0,\,\,\,\,\langle u_\alpha(\textbf{r})u_\beta(\textbf{r}')\rangle=(u^0_{\alpha})^2\delta_{\alpha\beta}\delta(\textbf{r}-\textbf{r}'),
\end{gather}
thus, the particles-impurity scattering can be characterized by the relaxation times $\tau_e^{-1}=m_e(u^0_e)^2$ and  $\tau_{ex}^{-1}=M(u^0_{ex})^2$ for the electrons and excitons, respectively, while $m_e$ and $M$ are their masses. 

For further progress, it is necessary to know the explicit forms of the nonlinear response functions $\mathbf{\Delta}_{ex}$ and $\mathbf{\Delta}_e$. 
They depend on the type of particle transport (ballistic or diffusive) and  the phase state of the exciton gas (normal or condensed). 
Below we will consistently analyze all these cases.

\section{\label{sec:level3}Coulomb drag of excitons in normal phase}

\begin{figure}[t!]
\includegraphics[width=0.9\linewidth]{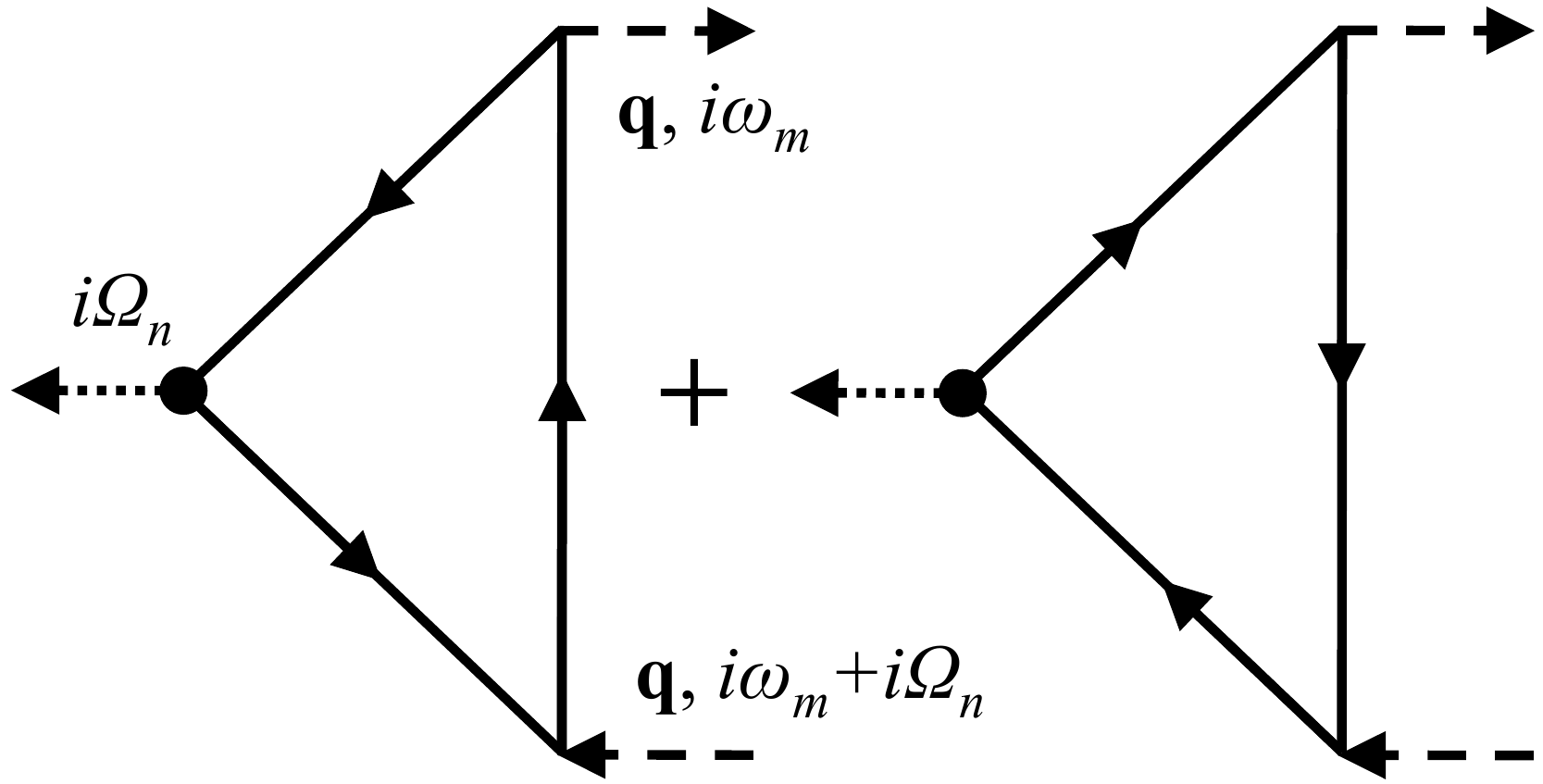}
\caption{The diagrammatic representation of the nonlinear response function \eqref{eq.I.7}. The solid lines correspond to the excitonic propagators $G_\textbf{p}(i\omega_{n})$, the dashed lines are for exciton-electron intraction potentials $V(\textbf{q},i\omega_m)$, and the filled circles depict the flux vertexes.}
\label{Fig2}
\end{figure}

In this section we consider temperatures higher than the critical temperature of the exciton BEC. 
Applying the Wick's theorem to~(\ref{eq.I.5b}) and performing the Fourier transform to the momentum space and the Matsubara frequency domain, we find
\begin{eqnarray}\label{eq.I.7}
&&\mathbf{\Delta}_{ex}(\textbf{q};i\Omega_n+i\omega_m,i\omega_m)
=\frac{g_{s}}{\beta}\sum_{\textbf{p},i\omega_{n'}}\, \frac{\textbf{p}}{M}\times
\\
\nonumber 
&&~~~\times\Bigl[
G_\textbf{p}(i\omega_{n'}-i\omega_m)
G_{\textbf{p}+\textbf{q}}(i\omega_{n'})
G_{\textbf{p}}(i\omega_{n'}-i\omega_m-i\Omega_n) \\
\nonumber
&&~~~~~~
+\{\textbf{q},i\omega_m,i\Omega_n ~~ \longrightarrow ~~ -\textbf{q},-i\omega_m,-i\Omega_n\}\Bigr],
\end{eqnarray}
where $G_\textbf{p}(i\omega_{n})$ is the excitonic propagator and $g_{s}=4$ is the spin degeneracy. Figure~\ref{Fig2} shows the graphic representation of Eq.~(\ref{eq.I.7}). The 2DEG response function has a similar structure.
%
%
%

\subsection{Diffusive regime}
\begin{figure}[t!]
\includegraphics[width=0.9\linewidth]{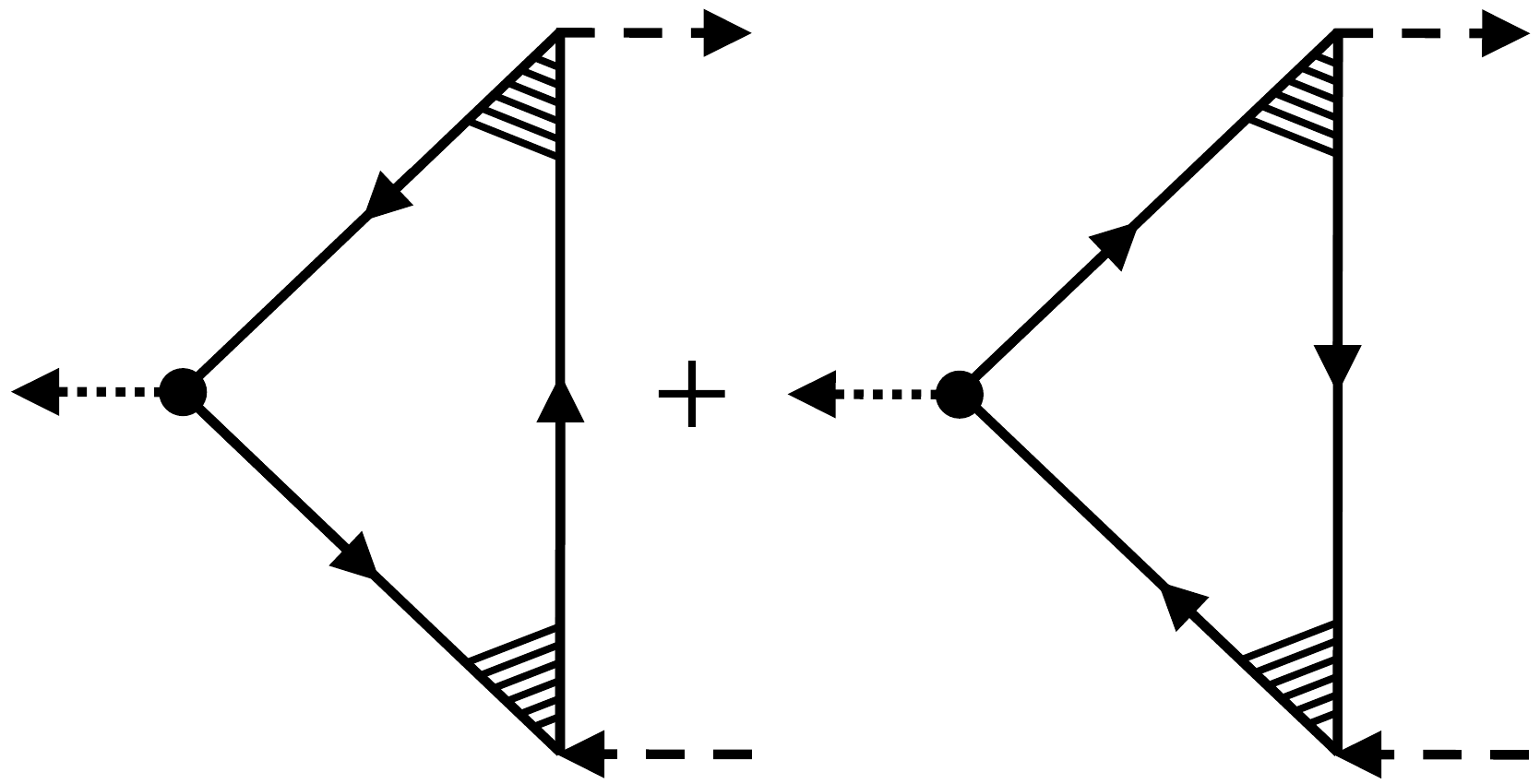}
\caption{ The diagrammatic representation of the nonlinear response function in the diffusive transport regime \eqref{eq.II.1}. The shading of the density vertexes $\Gamma_{\textbf{p}}(i\omega_n,i\omega_m)$ means the averaging over the random impurity field.}
\label{Fig3}
\end{figure}

In the diffusive regime of the exciton and electron motion, we assume  $\omega\tau_{e}\ll 1$, $ql_e\ll 1$ and $\omega\tau_{ex}\ll 1$, $ql_{ex}\ll 1$, where $l_e=v_F\tau_e$ and $l_{ex}=v_T\tau_{ex}$ are electron and exciton mean free pathes. 
Here $v_T=\sqrt{2T/M}$ is the mean value of exciton velocity determined by the temperature. 
To find the nonlinear response functions, we average the charge vertices, as it is shown in Fig.~\ref{Fig3}. 
The diagram implies that
\begin{eqnarray}\label{eq.II.1}
&&\mathbf{\Delta}_{ex}(\textbf{q};i\Omega_n+i\omega_m,i\omega_m)= g_s\sum_{\textbf{p},i\omega_{n'}}\frac{\textbf{p}}{M}\frac{1}{\beta}\times\, \\\nonumber
&&~\times \Bigl[ G_\textbf{p}(i\omega_{n'}-i\omega_m-i\Omega_n) \Gamma_{-\textbf{p}}(i\omega_{n'}-i\omega_m-i\Omega_n,i\omega_{n'})
\\\nonumber
&&~~~\times G_{\textbf{p}+\textbf{q}}(i\omega_{n'}) \Gamma_{\textbf{p}}(i\omega_{n'},i\omega_{n'}-i\omega_m)   G_{\textbf{p}}(i\omega_{n'}-i\omega_m)+
\\\nonumber
&&~~~+\{\textbf{q},i\omega_m,i\Omega_n\rightarrow -\textbf{q},-i\omega_m,-i\Omega_n\}\Bigr],
\end{eqnarray}
where $\Gamma_{\textbf{p}}$ is the density vertex averaged over impurity positions. It depends on the difference of the incoming and outgoing momenta and frequencies.

We assume that the exciton Green's functions are already averaged over the impurity potential. 
Switching from the summation to contour integrating and making the analytic continuations $i\omega_m+i\Omega_n\rightarrow\omega +i/2\tau_{ex}$ and $i\omega_m\rightarrow\omega -i/2\tau_{ex}$, we find

\begin{gather}
\nonumber
\mathbf{\Delta}_{ex}(\textbf{q};\omega^+,\omega^-)=g_s\omega\int \frac{d\varepsilon}{2\pi i}\, \frac{\partial n_B(\varepsilon)}{\partial\varepsilon} \sum_{\textbf{p}}\frac{\textbf{p}}{M}G_\textbf{p}^A(\varepsilon)G_\textbf{p}^R(\varepsilon)
\\\nonumber
\times\Bigl\{\Gamma_{\varepsilon}^{-}(\textbf{p},\omega)[G_{\textbf{p}+\textbf{q}}^R(\varepsilon+\omega)                      +G_{\textbf{p}-\textbf{q}}^A(\varepsilon-\omega)] -
\\
-\Gamma_{\varepsilon}^{+}(\textbf{p},\omega)[G_{\textbf{p}+\textbf{q}}^A(\varepsilon+\omega)    +G_{\textbf{p}-\textbf{q}}^R(\varepsilon-\omega)]\Bigr\},
\label{eq.II.3}
\end{gather}
where $$\Gamma_{\varepsilon}^{\pm}(\textbf{p},\omega)=\frac{\Theta(\varepsilon+\mu)}{\tau_{ex}(D_{\varepsilon+\mu} q^2\pm i\omega)},$$  $\mu$ is the chemical potential of excitons, $D_{\varepsilon+\mu}=\tau_{ex}v^2_{\varepsilon+\mu}/2$ is exciton diffusion coefficient at the mass shell $\varepsilon+\mu$, and $v$ is the exciton velocity.
In deriving~\eqref{eq.II.3}, we assumed $T\tau_{ex}\gg 1$, which means $T\gg 3$~K at $\tau_{ex}=2.5$~ps~\cite{Durnev2016}.  Together with $\omega\tau_{ex}\ll 1$, resulting in $\omega\ll T$,  it allows us to expand the distribution function over small $\omega/T$.

Since in the diffusive regime $\omega\ll1/\tau_{ex}$ and $q\ll1/l_{ex}$, we can also expand the Green's functions over small values of $\textbf{q}$ and $\omega$. Taking in~\eqref{eq.II.3} the integral over $\textbf{p}$, we come up with
\begin{gather}
\mathbf{\Delta}_{ex}(\textbf{q};\omega^+,\omega^-)= 4\omega\tau_{ex}\frac{\textbf{q}}{\pi}
\int\limits_0^\infty\frac{-D_\varepsilon q^2}{(D_\varepsilon q^2)^2+\omega^2}\frac{\partial n_B(\varepsilon-\mu)}{\partial\mu}d\varepsilon.
\label{eq.II.4}
\end{gather}

The nonlinear response function of the degenerate electron gas reads \cite{Kamenev_Oreg}
\begin{gather}\label{eq.II.5}
\mathbf{\Delta}_{e}(\textbf{q};\omega^-,\omega^+)=-\omega\tau_{e}\frac{\textbf{q}}{\pi}\frac{2Dq^2}{(Dq^2)^2+\omega^2},
\end{gather}
where $D=\tau_{e}v^2_{F}/2$ is the electron diffusion coefficient and $v_F$ is the Fermi velocity. 
Combining together~\eqref{eq.I.9},~\eqref{eq.II.4} and~\eqref{eq.II.5} and taking into account the screening of electron-exciton interaction (see Appendix~\ref{ApA}), we find
\begin{gather}
\nonumber
\mu_D=8\frac{e^5d^2\tau_e\tau_{ex}}{n_{ex}\epsilon^2\epsilon_T^2\pi}\frac{\partial}{\partial\mu} \int\limits_0^\infty \frac{d\varepsilon}{2\pi}\,  n_B(\varepsilon-\mu)\times\\
\nonumber
\times\int d\textbf{q}\,\int\limits_{-\infty}^\infty\frac{d\omega}{2\pi}\,\frac{\partial n_B(\omega)}{\partial\omega}\times \\
\times q^2\omega^2
\frac{e^{-2ql}Dq^2}{[(1+\kappa/q)^2(Dq^2)^2+\omega^2]}\frac{D_\varepsilon q^2}{[(D_\varepsilon q^2)^2+\omega^2]},
\label{eq.II.6}
\end{gather}
where $\kappa=2m_ee^2/\epsilon_0$ is the Thomas-Fermi momentum and $\epsilon_T$ is a static dielectric permittivity of exciton gas.
Performing the integrations in~\eqref{eq.II.6} in the limit $T\ll\varepsilon_F$ gives
\begin{gather}\label{eq.II.7}
\mu_D=-\frac{e}{2\pi n_{ex}}(\tau_{ex}T_c)(\kappa d)^2\left(\frac{\kappa}{p_F}\right)^2\alpha(\kappa l)\mathcal{F}\left(\frac{T}{T_c}\right),
\end{gather}
where
\begin{gather}\label{eq.II.8}
\mathcal{F}\left(x\right)=\frac{x\left(e^{1/x}-1\right)}{\Bigl[1+4\frac{M}{m_e}\kappa d(e^{1/x}-1)\Bigr]^2},\\\nonumber
\alpha(y)=\frac{1-4y-(2y)^2[1+(3+2y)e^{2y}\textrm{Ei}(-2y)]}{y^2}.
\end{gather}
Here $\textrm{Ei}(x)$ is the exponential integral function and $T_c=\pi n_{ex}/2M$ is the temperature of quantum degeneracy of the exciton gas. 

It is important to mention, that the inequality $\kappa l\gg 1$ often takes place. It result in simplification of Eq.~(\ref{eq.II.7}):
\begin{gather}\label{eq.II.9}
\mu_D=-\frac{3e}{4\pi n_{ex}}\frac{d^2\tau_{ex}T_c}{l^4p_F^2}\mathcal{F}\left(\frac{T}{T_c}\right).
\end{gather}
%
%

Note that negative sign of the cross-conductivity reflects that the excitons follow the direction of electron current, which flows in the direction opposite to the external field $\textbf{E}$.


\subsection{Quasi-ballistic regime}
In the quasi-ballistic regime ($q>1/l_{ex}$ or $\omega> 1/\tau_{ex}$), we can neglect the averaging of the density vertexes over the impurity field, putting $\Gamma_\textbf{q}(i\omega_n,i\omega_m)=1$ in Eq.~(\ref{eq.II.1}) but still averaging the propagators $G_\textbf{q}(i\omega_n)$. 
In this case the analytic expression for the nonlinear response function coincides with Eq.~(\ref{eq.I.7}). 
Switching from the Matsubara summation to integration along the real axis, we find: 
\begin{gather}
\nonumber
\mathbf{\Delta}_{ex}(\textbf{q};\omega^+,\omega^-)= \frac{2\tau_{ex}}{\pi}\int d\varepsilon\, \biggl\{ [n_B(\varepsilon+\omega)-n_B(\varepsilon)]
\\\nonumber
\times\sum_{\textbf{p}}\,\frac{\textbf{p}}{M}\Bigl(G_{\textbf{p}}^R(\varepsilon)-G_{\textbf{p}}^A(\varepsilon)\Bigr) \Bigl(G_{\textbf{p}+\textbf{q}}
^R(\varepsilon+\omega)-G_{\textbf{p}+\textbf{q}}^A(\varepsilon+\omega)\Bigr)
\\
+\{\textbf{q},\omega\longrightarrow -\textbf{q},-\omega\}\biggr\}.
\label{eq.III.1}
\end{gather}
The relation $g_{\textbf{p}}^R(\varepsilon)-g_{\textbf{p}}^A(\varepsilon)\simeq-2\pi i\delta(\varepsilon+\mu-\varepsilon_\textbf{p})$ allows us to carry out the integration over $\varepsilon$, yielding
\begin{gather}\label{eq.III.2}
\mathbf{\Delta}_{ex}(\textbf{q};\omega^+,\omega^-)=4\textbf{q}\frac{\tau_{ex}}{\pi}\int_0^\infty d\varepsilon_{\textbf{p}}\times
\\\nonumber
\times[n_B(\varepsilon_{\textbf{p}}-\mu+\omega)-n_B(\varepsilon_{\textbf{p}}-\mu)] \frac{\Theta(4\varepsilon_{\textbf{p}}\varepsilon_{\textbf{q}}-\omega^2)}{\sqrt{4\varepsilon_{\textbf{p}}\varepsilon_{\textbf{q}}-\omega^2}},
\end{gather}
where we replaced the integration variable from $\mathbf{p}$ to $\varepsilon_\mathbf{p}$; $\Theta(x)$ is the Heaviside step function, which imposes the upper limit on the integration domain $\omega<q\sqrt{2\varepsilon_\mathbf{p}/M}$. We also note that the exponential function from Eq.~\eqref{eq.I.10} limits the value of the momentum, $q<q_{max}\sim 1/l$, while the distribution functions in Eq.~(\ref{eq.III.2}) define $\varepsilon_{\mathbf{p},max}\sim T$.
Denoting $\omega_{max}=q_{max}\sqrt{2\varepsilon_{\mathbf{p},max}/M}$ one finds 
$\omega_{max}/T\sim2/p_Tl$, where $p_T=Mv_T$.  
If $p_Tl/2\gg 1$,  which holds for high $T$ or large  $l$, one can expand the distribution functions in Eq.~(\ref{eq.III.2}) in series over $\omega/T$. 
Using the expression for the nonlinear response function for quasiballistic electrons \cite{Kamenev_Oreg}
\begin{gather}\label{eq.III.3}
\mathbf{\Delta}_{e}(\textbf{q};\omega^-,\omega^+)=-\frac{2D\textbf{q}}{\varepsilon_F}\frac{m}{\pi}\frac{\omega}{v_Fq} \Theta[(v_Fq)^2-\omega^2],
\end{gather}
we find the cross-mobility in the quasiballistic regime:
\begin{gather}\label{eq.III.4}
\mu_D=-8T\frac{e^5d^2\tau_e\tau_{ex}}{n_{ex}\epsilon^2\epsilon_T^2\pi^2v_F}\int\limits_0^\infty dq\, \frac{q^4e^{-2ql}}{(q+\kappa)^2} \int\limits_0^\infty d\omega
\\\nonumber
\times\frac{\partial}{\partial\mu}\int\limits_0^\infty d\varepsilon_p\, n_B(\varepsilon_{\textbf{p}}-\mu) \frac{\Theta(4\varepsilon_{\textbf{p}}\varepsilon_{\textbf{q}}-\omega^2)} {\sqrt{4\varepsilon_{\textbf{p}}\varepsilon_{\textbf{q}}-\omega^2}}.
\end{gather}
The direct calculation of the integrals in~\eqref{eq.III.4} gives
\begin{gather}\label{eq.III.4a}
\mu_D=-\frac{e}{n_{ex}\pi}(T_c\tau_{ex})(T_c\tau_e)(\kappa d)^2\left(\frac{\kappa}{p_F}\right)^2
\\\nonumber\times\left(\frac{p_F\kappa}{mT_c}\right)\mathcal{F}\left(\frac{T}{T_c}\right)\beta(\kappa l),~\textrm{where}
\\
\nonumber
\beta(y)=\frac{1-2y+\frac{3}{2}(2y)^2+\frac{(2y)^3}{2}\left[1+2(2+y)e^{2y}\textrm{Ei}(-2y)\right]}{y^3}.
\end{gather}
Similar to the diffusive case, Eq.~\eqref{eq.III.4} essentially simplifies if $\kappa l\gg 1$:
\begin{gather}\label{eq.III.5}
\mu_D=-\frac{3e}{\pi n_{ex}} \frac{d^2\tau_e\tau_{ex}T_c}{l^5m_ep_F}\mathcal{F}\left(\frac{T}{T_c}\right).
\end{gather}
%


\section{\label{sec:level4}Coulomb drag of excitons in presence of the BEC phase}

If the temperature is lower than the critical temperature of a BEC formation, the ground state of the excitonic subsystem becomes macroscopically occupied. It constitutes the Bose-Einstain condensation phenomenon.
We will assume the temperature to be low enough to consider the condensate density nearly equal to the one at zero temperature: $(n_c(0)-n_c(T))/n_c(0)\ll 1$. This assumption allows us putting $T=0$ in the evaluation of the BEC response function and assuming $n_c(0)\equiv n_c=n_{ex}$.

Furthermore, we will represent the exciton field operators as a sum of two terms,
\begin{gather}\label{eq.IV.1}
\hat{\Psi}(\textbf{R},t)= \hat{\xi}_0+\hat{\varphi}(\textbf{R},t), 
\\\nonumber
\hat{\Psi}^+(\textbf{R},t)= \hat{\xi}_0^++\hat{\varphi}^+(\textbf{R},t),
\end{gather}
where $\hat{\xi}_0$ and $\hat{\xi}_0^{+}$ ($\hat{\varphi}$ and $\hat{\varphi}^{+}$) are the annihilation and creation operators of an exciton in the ground (excited) state, and the full exciton density operator reads $\hat{n}=\hat{\Psi}^+ \hat{\Psi}$.
Macroscopically large occupation of the $\textbf{p}=0$ state enables a conventional replacement $\hat{\xi}_0, \hat{\xi}_0^+\rightarrow \sqrt{n_c}$
.
Then Eq.~(\ref{eq.I.5b}) transforms into
\begin{gather}\label{eq.IV.2}
\mathbf{\Delta}_{c}=-g_sn_c\langle T_\tau\hat{\textbf{J}} (\textbf{R},\tau)\Bigl[ 
\hat{\varphi}(x_1)\hat{\varphi}(x_2)+
\hat{\varphi}(x_1)\hat{\varphi}^+(x_2)+
\\\nonumber
+\hat{\varphi}^+(x_1)\hat{\varphi}(x_2)+
\hat{\varphi}^+(x_1)\hat{\varphi}^+(x_2)
\Bigr]\rangle,
\end{gather}
where $$\hat{\textbf{J}} (\textbf{R},\tau)=\frac{1}{2Mi}\lim_{\textbf{R}'\rightarrow\textbf{R}}(\nabla_{\textbf{R}'}-\nabla_{\textbf{R}}) \hat{\varphi}^+(\textbf{R}',\tau) \hat{\varphi}(\textbf{R},\tau)$$
is an operator of condensed excitons flux, the factor $g_s$ in~\eqref{eq.IV.2} is due to the spin degeneracy, and a short-hand notation $x=(\textbf{R},\tau)$ is used. 
We note, that Eq.~\eqref{eq.IV.2} gives a term proportional to $n_c^2$, which does not contribute to the flux. 
Furthermore, in deriving Eq.~(\ref{eq.IV.2}) we neglected the term proportional to $\hat{\varphi}(x_1)\hat{\varphi}^+(x_1) \hat{\varphi}(x_2)\hat{\varphi}^+(x_2)$, which describes the drag of non-condensed particles.

\begin{figure}[t!]
\includegraphics[width=0.45\linewidth]{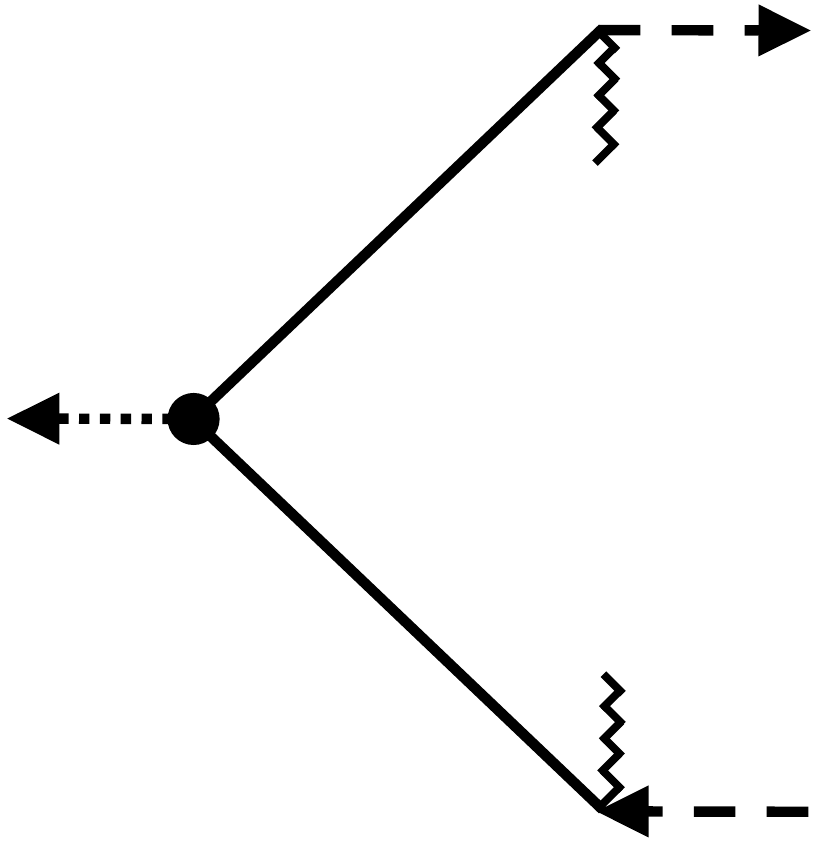}
\caption{The structure of each term in the expansion of the condensate nonlinear response function~\eqref{eq.IV.2} by the Wick's theorem. The solid lines are for any propagators \eqref{eq.IV.3a}, while the wavy lines denote $\sqrt{n_c}$.} 
\label{Fig4}
\end{figure}

Next, the Wick's theorem allows us to rewrite Eq.~(\ref{eq.IV.2}) as a sum of products of the normal and anomalous propagators, defined as
\begin{gather} \label{eq.IV.3a}
\left(
             \begin{array}{cc}
               \mathfrak{G}(x-x') & \mathfrak{F}(x-x') \\
               \mathfrak{F}^+(x-x') & \tilde{\mathfrak{G}}(x-x') \\
             \end{array}
                   \right)=\\\nonumber=
\left(
           \begin{array}{cc}
             -\langle T_\tau\varphi(x)\varphi^+(x')\rangle & -\langle T_\tau\varphi(x)\varphi(x')\rangle \\
             -\langle T_\tau\varphi^+(x)\varphi^+(x')\rangle & -\langle T_\tau\varphi^+(x)\varphi(x')\rangle \\
            \end{array}
                  \right).
\end{gather}
The terms in that sum have an identical structure, shown in Fig.~\ref{Fig4}. Switching to the momentum and the Matsubara frequency domains, after some algebra we find
\begin{eqnarray} \label{eq.IV.3}
&&\mathbf{\Delta}_{c}(\textbf{q};i\Omega_n+i\omega_m,i\omega_m)
\\\nonumber
&&~~~=-\frac{g_sn_c\textbf{q}}{M}\biggl[
    \Bigl(\mathfrak{G}_\textbf{q}(i\omega_m) +\mathfrak{F}_\textbf{-q}(-i\omega_m)\Bigr)
    \\\nonumber
  &&~~~~~  \times\Bigl(\mathfrak{G}_\textbf{q}(i\Omega_n+i\omega_m) +\mathfrak{F}_\textbf{-q}(-i\Omega_n-i\omega_m)\Bigr)
    \\\nonumber
 &&~~~~~~~~~   -\{\textbf{q},i\omega_m,i\Omega_n\rightarrow -\textbf{q},-i\omega_m,-i\Omega_n\}
\biggr],
\end{eqnarray}
where we assumed $\mathfrak{F}^+=\mathfrak{F}$.
Note that Eq.~(\ref{eq.IV.3})  includes neither momentum nor energy summations since the condensate propagators have no coordinate or time dependencies
(for example, $\langle\hat{\xi}_0\hat{\xi}_0\rangle=n_c$). 

Within the dilute Bose gas model, the propagators in~\eqref{eq.IV.3} read \cite{Capogrosso}
\begin{gather} \label{eq.IV.4}
\mathfrak{G}_\textbf{q}(i\omega_m)=
\frac{i\omega_m+\varepsilon_\textbf{q}+n_cg} {(i\omega_m)^2-\omega_\textbf{q}^2},
\\\nonumber
\mathfrak{F}_\textbf{q}(i\omega_m)=
\frac{-n_cg} {(i\omega_m)^2-\omega_\textbf{q}^2},
\end{gather}
where $\omega_\textbf{q}=\sqrt{\varepsilon_\textbf{q}(2gn_c+\varepsilon_\textbf{q})}=sq\sqrt{1+(q\xi)^2}$ is a Bogoliubov quasi-particle dispersion, $s=\sqrt{gn_c/M}$ is their phase velocity, $\xi=1/(2Ms^2)$ is a healing length, and the exciton-exciton interaction strength estimates as $g=4\pi e^2d/\varepsilon$. 
To derive an expression for $\mathbf{\Delta}_{c}(\textbf{q};\omega^+,\omega^-)$, we should perform the analytic continuations in Eq.~(\ref{eq.IV.3}), similar to one mentioned right before Eq.~(\ref{eq.II.3}): $i\omega_m+i\Omega_n\rightarrow\omega +i\gamma_q$ and $i\omega_m\rightarrow\omega -i\gamma_q$. 
Clearly, the difference is only in the imaginary term $i\gamma_q$, responsible to the scattering on impurities.

The influence of the impurity potential on transport of excitons and exciton polaritons in presence of the BEC has been analyzed in Refs.~\onlinecite{KCh3, KS}. The scattering of the BEC particles via impurities results in a damping of Bogoliubov quasiparticles in the linear domain ($q\xi\ll 1$) of their dispersion: $\omega_q=sq\rightarrow\omega_q-i\gamma_q$, where $\gamma_q=(q\xi)^3/\tau_{ex}$ and $\tau_{ex}$ is the (normal phase) exciton-impurity scattering time. Since in our paper we consider the case $\tau_e\sim\tau_{ex}$, we yield $\gamma_q\ll 1/\tau_e$ and the exction motion we can consider as purely ballistic. 

Substituting Eq.~(\ref{eq.IV.4}) in Eq.~(\ref{eq.IV.3}) and performing analytic continuation, we obtain
\begin{gather} \label{eq.IV.5}
\mathbf{\Delta}_{c}(\textbf{q};\omega^+,\omega^-)=
\frac{-4g_sn_c\textbf{q}\omega\varepsilon_\textbf{q}} {M[(\omega-\omega_\textbf{q})^2+\gamma_q^2][(\omega+\omega_\textbf{q})^2+\gamma_q^2]}. 
\end{gather}

The transport in the 2DEG subsystem can be either diffusive or ballistic. Let us consider them separately.


\subsection{Diffusive regime}
Taking into account the screening effect (see Appendix~\ref{ApA}), we arrive at the expression for the cross-mobility by the substitution of Eq.~(\ref{eq.II.5}) and (\ref{eq.IV.5}) in the general formula~\eqref{eq.I.9} and by utilizing the replacements $q\rightarrow x/2l$ and $\omega\rightarrow sy/2l$:
%
\begin{gather} \label{eq.IV.6}
\mu_D=-\frac{2e}{\pi}\frac{e^4d^2}{\epsilon^2} \frac{\tau_e}{TM^2s(2l)^3D} 
\\\nonumber
\times \int\limits_{-\infty}^\infty dy 
y^2 \textrm{sh}^{-2}\left( \frac{sy}{4lT} \right)
\int\limits_{0}^\infty dx\frac{x^7e^{-x}}{|A(x,y)|^2},
\end{gather}
where 
\begin{gather} \label{eq.IV.7}
A(x,y)=2d\kappa x^4e^{-x}
\\\nonumber
+\left[x(x+2l\kappa)-iy\frac{2ls}{D}\right] \left[\left(y+i\frac{x^3\xi^3}{s(2l)^2\tau_{ex}}\right)^2-x^2\right]
.
\end{gather}
The integrand represents an extremely peaked function, when its argument approaches $y=\pm xb$ (where $b=\sqrt{(x+2l\kappa-2d\kappa xe^{-x})/(x+2l\kappa)}$).
Thus the integration over $y$ can be performed, yielding
\begin{gather} \label{eq.IV.8}
\mu_D=-8e\frac{e^4d^2}{\epsilon^2} \frac{\tau_{ex}Ms^3}{Tlv_F^2} 
\\\nonumber
\times\int\limits_{0}^\infty dx\frac{x^2e^{-x}}{(x+2l\kappa)^2}
\textrm{sh}^{-2}\left( \frac{sbx}{4lT} \right).
\end{gather}
The integral in~\eqref{eq.IV.8} cannot be taken analytically. However, the typical values of experimental parameters suggest that $l\kappa\gg 1$ and we can neglect $x$ in the denominator. 
For instance, in GaAs-based alloys, $1/\kappa\simeq 5 ~\textrm{nm} \ll 50 \textrm{nm} <l$. 
Then the integration in Eq.~\eqref{eq.IV.8} can be done
%
\begin{gather} \label{eq.IV.9}
\mu_D=-2^7e\frac{e^4d^2}{\epsilon^2} \frac{\tau_{ex}MlT^3}{v_F^2\kappa^2s}\times 
\\\nonumber
\times\left\{ \frac{s}{lT}\psi_1\left(\frac{2lT}{s}\right) +\psi_2\left(\frac{2lT}{s} \right) \right\},
\end{gather}
where $\psi_n(x)$ is the polygamma function and we took $b=1$. 

The argument of $\psi_{1,2}$ functions is a ratio of the thermal energy $T$ to the maximal energy of a Bogoliubov quasiparticle, $\omega_{max}=sq_{max}=s/2l$, being excited by the electric current. 
Thus it is appropriate to consider two limiting cases. When the temperature is much lower than the Bogoliubov energy quantum, $T\ll \omega_{max}$,  we have 
$$\frac{s}{lT}\psi_1\left(\frac{2lT}{s}\right) +\psi_2\left(\frac{2lT}{s} \right)\simeq \frac{\pi^2s}{6lT},$$ and 
the cross-mobility reads
\begin{gather} \label{eq.IV.10}
\mu_D\simeq-\frac{2^6\pi^2e}{3} \frac{e^4d^2}{\epsilon^2} \frac{\tau_{ex}MT^2}{v_F^2\kappa^2}=
\\\nonumber
=-\frac{16\pi^2e}{3}  \frac{d^2\tau_{ex}MT^2}{p_F^2}.
\end{gather}
In the opposite case, $T\gg \omega_{max}$, the expression in the curly brackets in~\eqref{eq.IV.9} becomes
\begin{gather}
\nonumber
\frac{s}{lT}\psi_1\left(\frac{2lT}{s}\right) +\psi_2\left(\frac{2lT}{s} \right)=
\\\nonumber
=\left(\frac{s}{2lT}\right)^2-\frac{1}{6}\left(\frac{s}{2lT}\right)^4
+O\left[\left(\frac{s}{2lT}\right)^5\right],
\end{gather}
and Eq.~\eqref{eq.IV.9} takes the form
\begin{gather} \label{eq.IV.10a}
\mu_D\simeq-8e  \frac{d^2\tau_{ex}MsT}{p_F^2l} \left(1-\frac{s^2}{24l^2T^2} \right).
\end{gather}
We keep the second term of the expansion here since the ratio $s/2lT$ cannot be much smaller than unity at low temperatures. For instance, the estimations give $s\simeq 2*10^6 ~\textrm{cm/s}$ at $n_c=10^{10} ~\textrm{cm}^{-2}$, thus $s/2lT\simeq 0.85$ at $l=200~\textrm{nm}$ and $T=0.5~\textrm{K}$. 
However, the two terms in~\eqref{eq.IV.10a} give a good enough agreement with~\eqref{eq.IV.9} at $T>0.5$~K (see Fig.~\ref{Fig_Dif_T0}), thus we disregard the higher-order terms.

\begin{figure}[t!]
\includegraphics[width=0.99\linewidth]{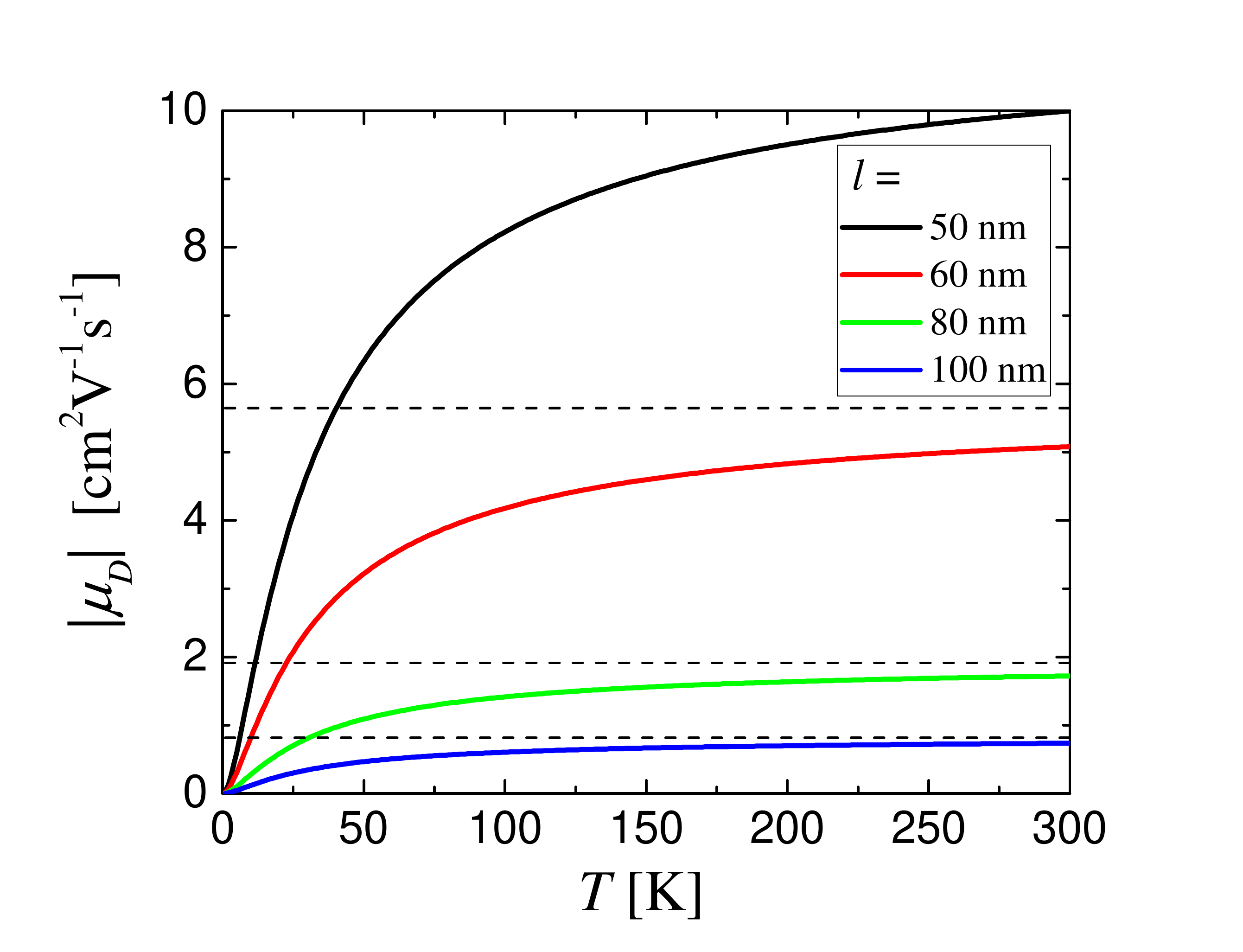}
\includegraphics[width=0.99\linewidth]{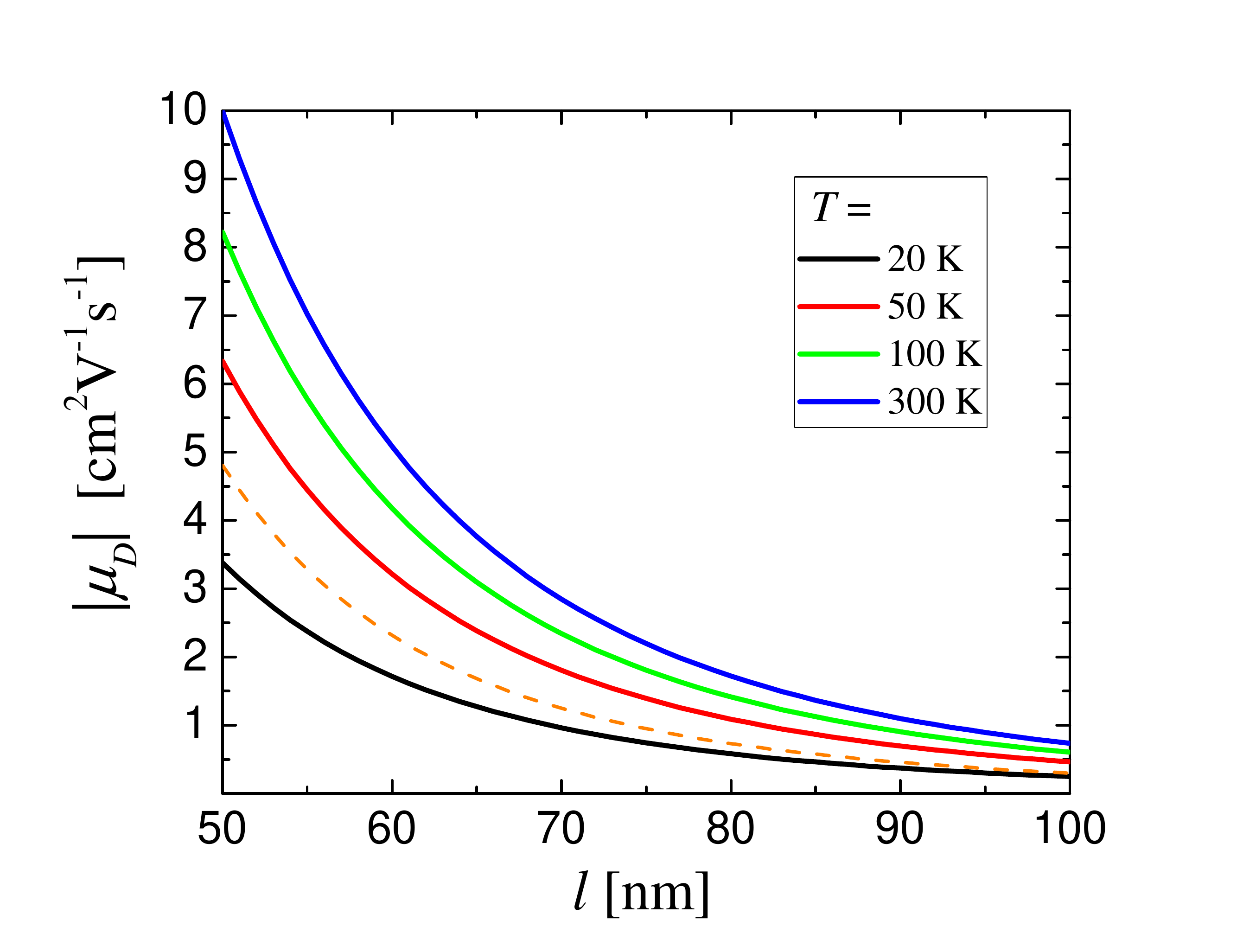}
\caption{
The cross--mobility (by the absolute value) as a function of temperature (upper panel) and the interlayer distance (lower panel) in the normal state of the exciton gas and in the regime of diffusive motion of electrons.
Dashed black lines stand for the high-temperature asymptotics.
Dashed orange curve depicts Eqs.~\eqref{eq.II.9} at $T=20$~K.
} 
\label{Fig_Dif_T}
\end{figure}
\begin{figure}[t!]
\includegraphics[width=0.99\linewidth]{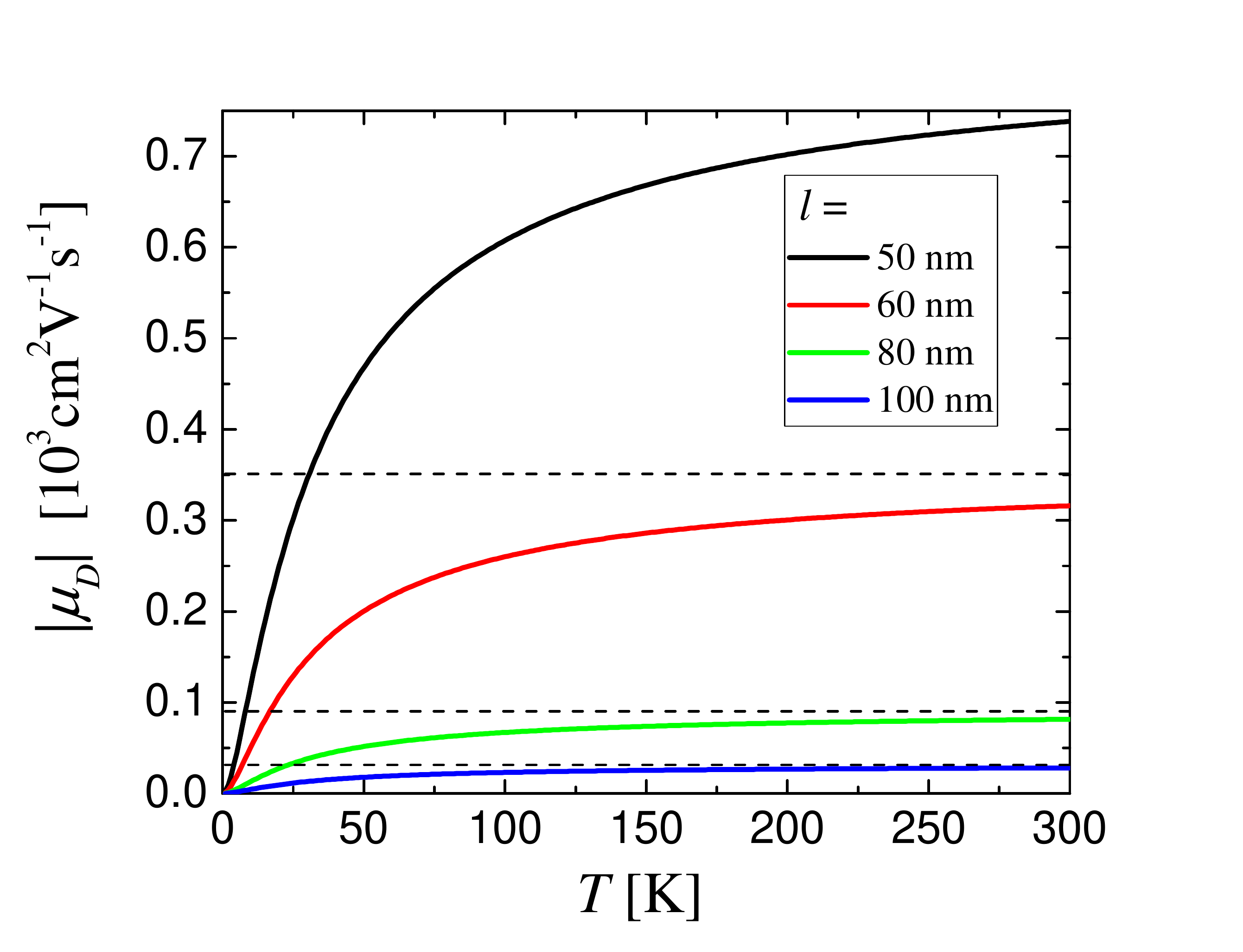}
\includegraphics[width=0.99\linewidth]{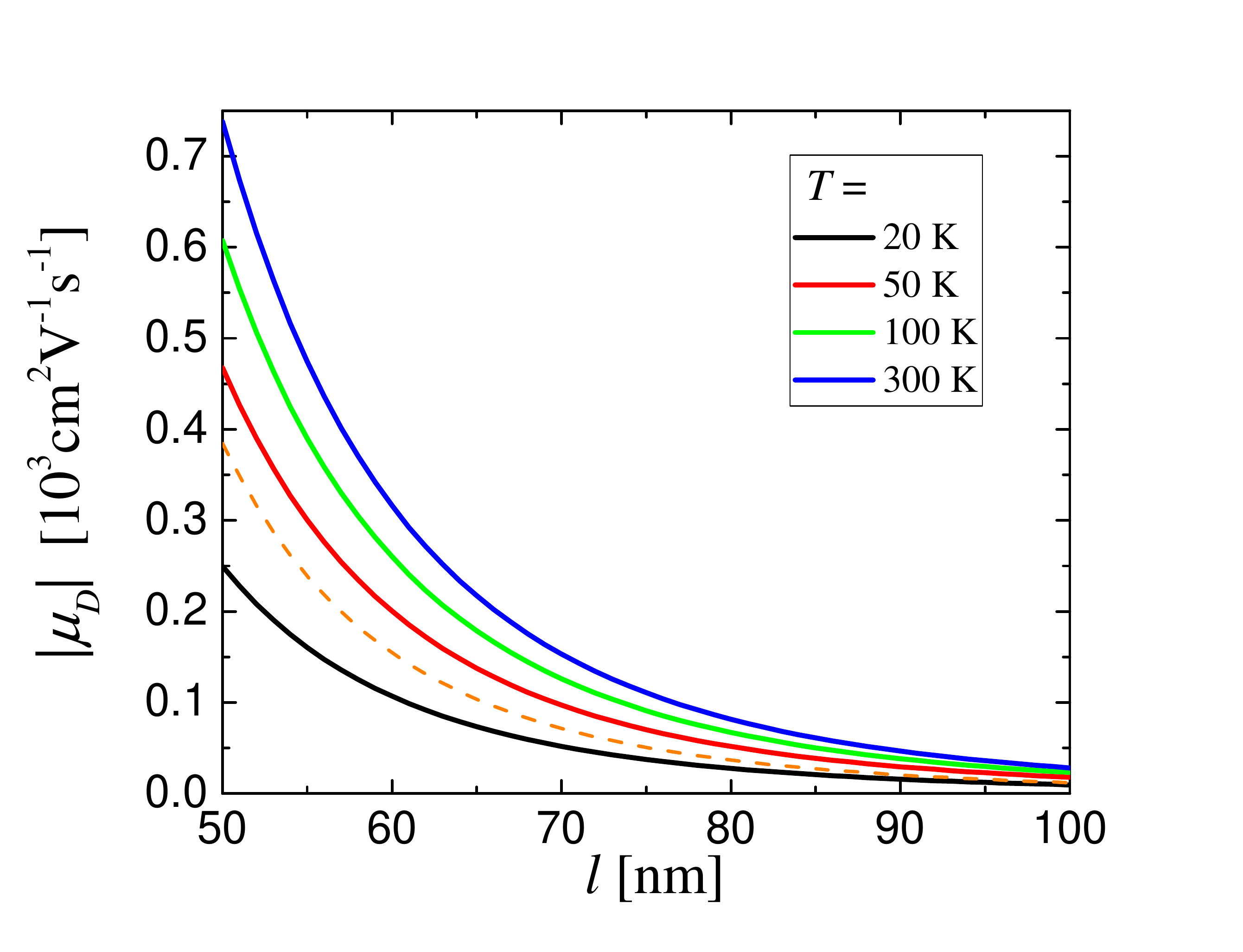}
\caption{
The same as in Fig.~\ref{Fig_Dif_T} for ballistic regime.  The orange dashed line in lower panel depicts Eq.~\eqref{eq.III.5} at $20$~K.
} 
\label{Fig_Bal_T}
\end{figure}
%
%
%


\subsection{Quasi-ballistic regime}
Utilizing Eq.~(\ref{eq.III.3}) instead of Eq.~(\ref{eq.II.5}) in the ballistic case we arrive at
\begin{gather} \label{eq.IV.11}
\mu_D=-\frac{2e}{\pi}\frac{e^4d^2}{\epsilon^2} \frac{\tau_e}{TM^2v_Fs(2l)^4} 
\\\nonumber
\times \int\limits_{-\infty}^\infty dy 
y^2 \textrm{sh}^{-2}\left( \frac{sy}{4lT} \right)
\int\limits_{0}^\infty dx\frac{x^6e^{-x}\Theta[(v_Fx/s)^2-y^2]}{|B(x,y)|^2},
\end{gather}
where 
\begin{gather} \label{eq.IV.12}
B(x,y)=
\\\nonumber
(x+2l\kappa) \left[\left(y+i\frac{x^3\xi^3}{s(2l)^2\tau_{ex}}\right)^2-x^2\right]
+2d\kappa x^3e^{-x}.
\end{gather}
The integrand in Eq.~(\ref{eq.IV.11}) is peaked as well as in Eq.~(\ref{eq.IV.6}), hence the same approach can be done. 
\begin{figure}[t!]
\includegraphics[width=0.99\linewidth]{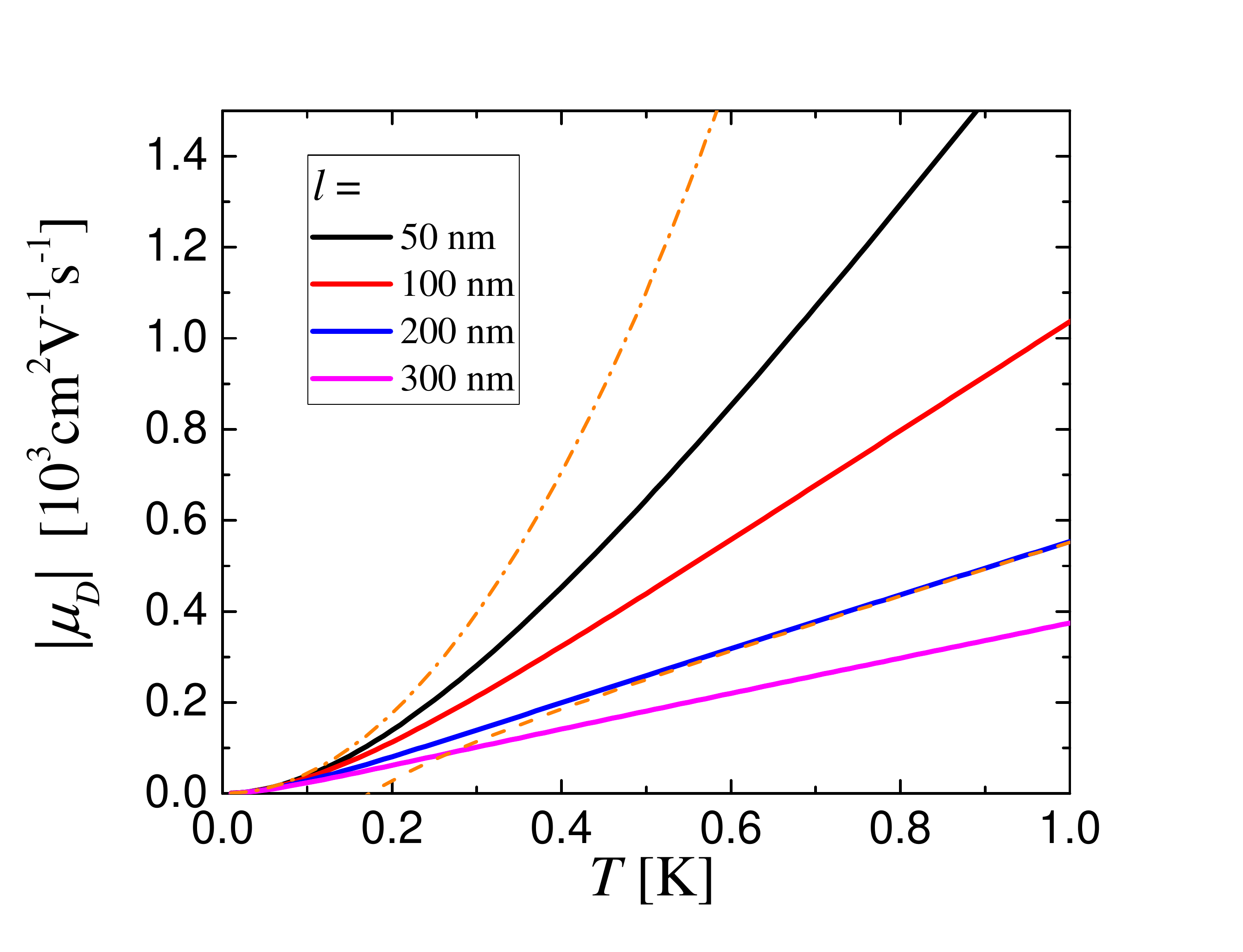}
\includegraphics[width=0.99\linewidth]{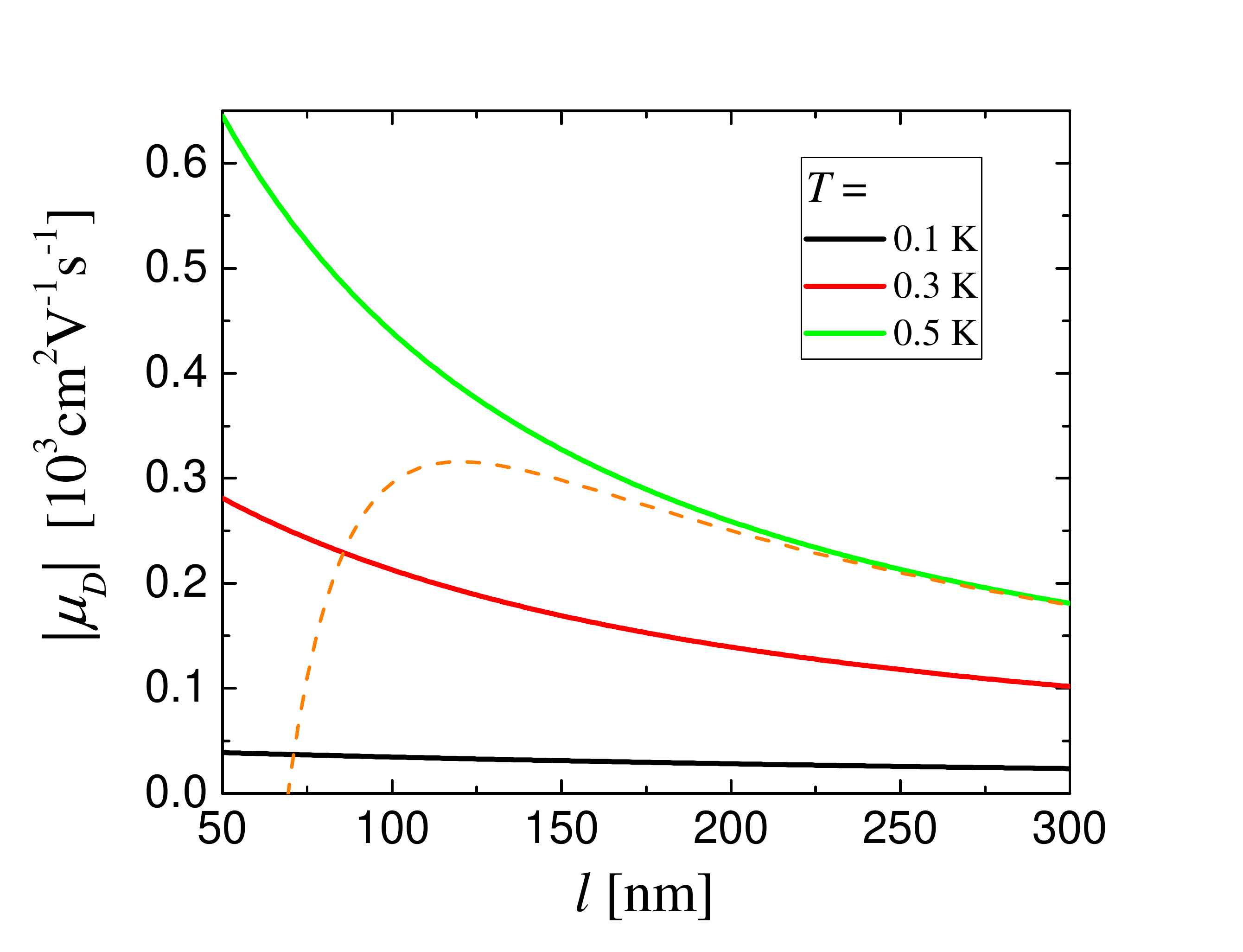}
\caption{
The cross--mobility (by the absolute value) as a function of temperature (upper panel) and the interlayer distance (lower panel) in the regime of excitonic condensation and the regime of diffusive motion of electrons. 
The orange dash-dotted curve depicts Eq.~\eqref{eq.IV.10} and stands for the low-temperature asymptotics. 
The dashed curve in the upper panel depicts~\eqref{eq.IV.10a} at $l=200$~nm and shows the high-temperature asymptotics. 
The lower-panel orange dashed curve depicts~\eqref{eq.IV.10a} for $T=0.5$~K and shows the asymptotics for a high value of $l$.
}
\label{Fig_Dif_T0}
\end{figure}
\begin{figure}[t!]
\includegraphics[width=0.99\linewidth]{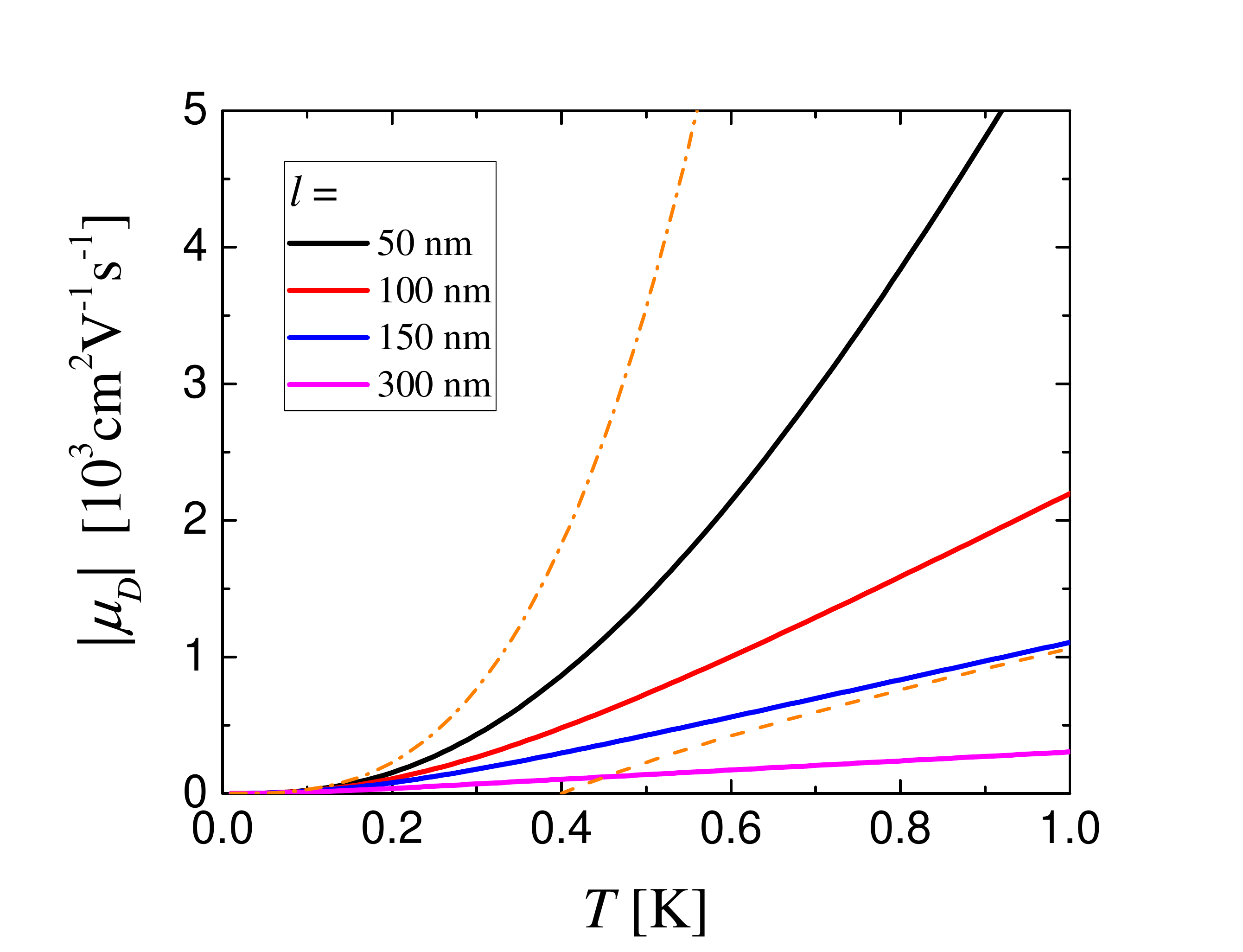}
\includegraphics[width=0.99\linewidth]{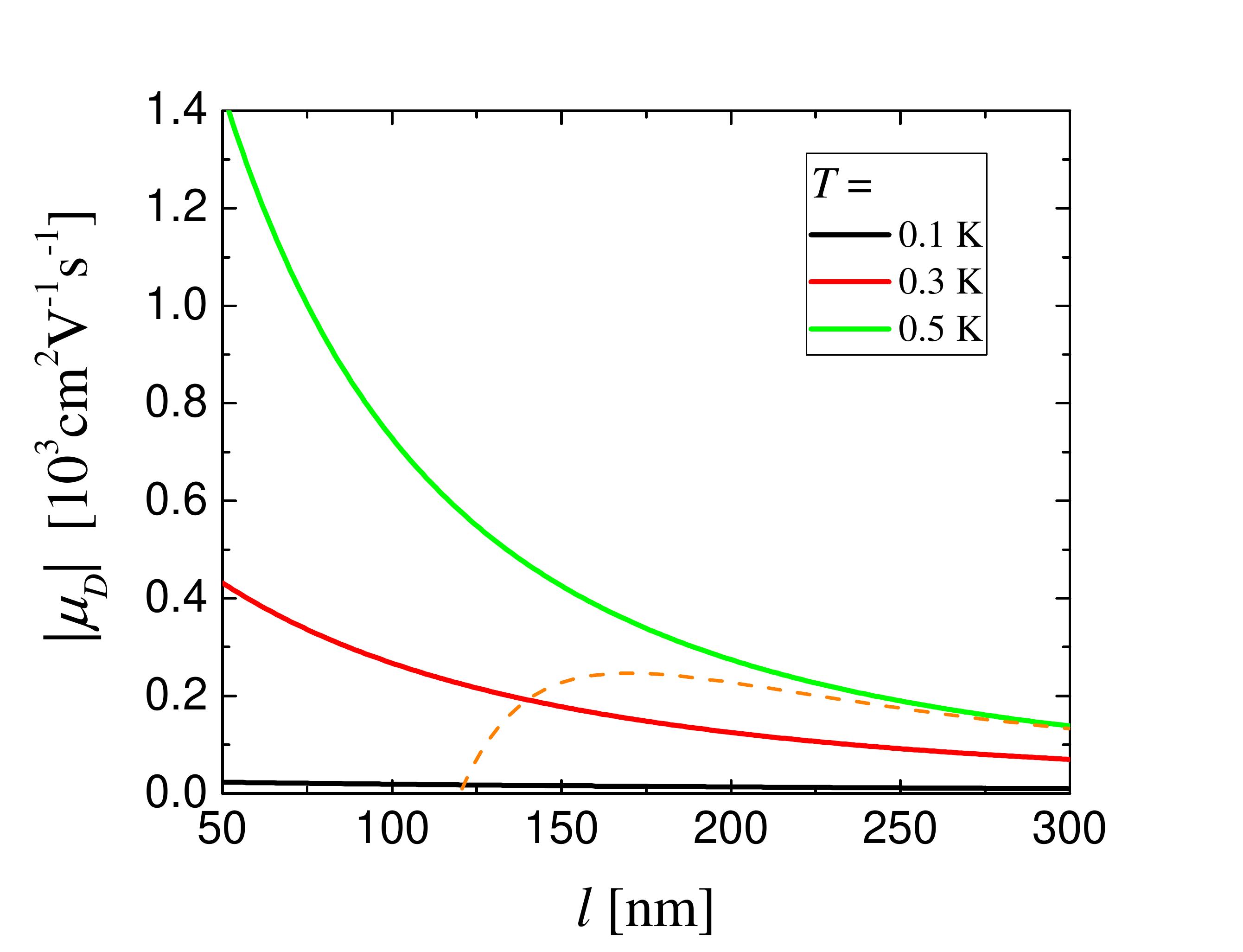}
\caption{
The same as in Fig.~\ref{Fig_Dif_T0} for the ballistic regime of electron motion. 
The orange dash-dotted curve depicts Eq.~\eqref{eq.IV.15} and stands for the low-temperature asymptotics. 
The dashed curve in the upper panel depicts~\eqref{eq.IV.16} at $l=150$~nm and shows the high-temperature asymptotics. 
The lower-panel orange dashed curve depicts~\eqref{eq.IV.16} for $T=0.5$~K and shows the asymptotics for a high value of $l$.
} 
\label{Fig_Bal_T0}
\end{figure}
Note, $b<v_F/s$ at reasonable parameters, thus we can suppose $\Theta[(v_Fx/s)^2-y^2]=1$. It gives
\begin{gather} \label{eq.IV.13}
\mu_D=-8e\frac{e^4d^2}{\epsilon^2} \frac{\tau_e\tau_{ex}Ms^3}{Tv_F(2l)^2} 
\\\nonumber
\times\int\limits_{0}^\infty dx\frac{x^3e^{-x}}{(x+2l\kappa)^2}
\textrm{sh}^{-2}\left( \frac{sbx}{4lT} \right).
\end{gather}
Taking this integral with the same assumptions like those which we utilized to find~\eqref{eq.IV.8}, we obtain
\begin{gather} \label{eq.IV.14}
\mu_D=16e \frac{d^2\tau_{e}\tau_{ex}MlT^4} {m_es^2p_F} \times
\\\nonumber
\times\left\{ 
\frac{3s}{2lT}\psi_2\left(\frac{2lT}{s}\right)+\psi_3\left(\frac{2lT}{s}\right)
\right\}.
\end{gather}
%
In the case $T\ll \omega_{max}$ the relation
$$\frac{3s}{2lT}\psi_2\left(\frac{2lT}{s}\right)+\psi_3\left(\frac{2lT}{s}\right)
\simeq -3\zeta(3)\frac{s}{lT}$$
takes place, and thus
\begin{gather} \label{eq.IV.15}
\mu_D\simeq-48\zeta(3)e \frac{d^2\tau_{e}\tau_{ex}MT^3}{m_esp_F},
\end{gather}
where $\zeta$ is the Riemann zeta-function.
On the contrary, if $T\gg \omega_{max}$, then
\begin{gather}
\nonumber
\frac{3s}{2lT}\psi_2\left(\frac{2lT}{s}\right)+\psi_3\left(\frac{2lT}{s}\right)=
\\\nonumber
=-\left(\frac{s}{2lT}\right)^3+ \frac{1}{2} \left(\frac{s}{2lT}\right)^5
+O\left[\left(\frac{s}{2lT}\right)^6\right],
\end{gather}
which allows us to write the expression for cross-mobility in the form
\begin{gather} \label{eq.IV.16}
\mu_D\simeq-2e  \frac{d^2\tau_{e}\tau_{ex}MsT} {m_ep_Fl^2} \left(1-\frac{s^2}{8l^2T^2} \right).
\end{gather}
%


\section*{Discussion}
Let us compare the cross-mobilities in different regimes for various temperatures and interlayer distances (separations between the electronic and excitonic layers). We will use typical parameters of GaAs-based heterostructures: $m_e=0.067m_0$, $M=0.5m_0$ (where $m_0$ is a free electron mass), $\epsilon=12.9$, $d=10$~nm, $n_{ex}=10^{10}$~cm$^{-2}$, and $v_F=10^7$~cm$/$s.

We begin with considering the diffusive regime without the excitonic condensate. 
Note that Eq.~\eqref{eq.II.7} does not explicitly include the electron relaxation time, resulting from our assumption that the diffusive constant of electrons is large in comparison with the one of excitons. 
However, this time is implicitly restricted by the condition that in the diffusive regime $l_e\ll l $ so $\tau_e\ll l/v_F=0.5$~ps at $l=50$~nm.

Figure~\ref{Fig_Dif_T} shows the dependence of the cross-mobility on temperature and the inter-layer spacing $l$. 
At high temperatures ($x\gg 1$ in \eqref{eq.II.8}), we have $\mathcal{F}\left(x\right)\approx 1$  and the cross-mobility \eqref{eq.II.7} does not depend on temperature, whereas at low temperatures ($x\ll 1$) it decreases to zero exponentially, $\mathcal{F}\left(x\right)\sim xe^{-1/x}$.

In the ballistic regime, the condition $\tau_{ex}v_T>l$ restricts the domain of~\eqref{eq.III.4a} validity. 
At $l=100$~nm and $\tau_{ex}=2.5$~ps~\cite{Durnev2016} the temperature must be higher than $T_{b}=Mv_T^2/2\sim 25$~K. Then the inequality $p_Tl/2\gg 1$ is also satisfied.
At lower temperatures, the motion of excitons should be treated diffusive, while the motion of electrons will remain ballistic (we  do not consider this case).
Figure~\ref{Fig_Bal_T} shows  qualitatively the same temperature dependence of $\mu_D$ like in the diffusive case. 
We remind, however, that there exists a boundary value of temperature $T_{b}$, below which our treatment is not applicable.
Note, that in the ballistic case the cross-mobility is two orders of magnitude higher than in the diffusive regime, where the friction due to the impurity field is strong.

Figures~\ref{Fig_Dif_T0} and~\ref{Fig_Bal_T0} (upper panels) show an astonishingly high increase of the absolute values of $\mu_D$ with temperature, which is expectable at low temperatures. 
Comparing these figures with Figs.~\ref{Fig_Dif_T} and~\ref{Fig_Bal_T} (upper panels), we conclude that the cross-mobility depends on temperature non-monotonously. 
Indeed, when the condensation takes place, $\mu_D$ should experience a substantial increase (by the absolute value).
From the physical point of view, the reason for such increase  consists in an extremely weak attenuation of the Bogoliubov quasiparticles due to the scattering on impurity ions, as opposed to a relatively large damping of excitons in the normal phase.
%



\section*{Conclusions}
We have presented a theory of the Coulomb drag effect in a hybrid Bose-Fermi system consisting of spatially separated two-dimensional electron and indirect exciton gases, considering both the normal and condensate phases of the excitonic subsystem. 
We have calculated the cross-mobility in the system and studied its temperature dependence for different spacings between the electron and exciton layers. 
We conclude that the inter-exciton interaction strongly influences the cross-mobility.
In the normal phase of the exciton gas, this interaction determines the temperature dependence of cross-mobility of the electron-exciton mixture and results in its exponential damping with the decrease of 
temperature. 
At low temperatures, the inter-exciton interaction leads to a considerable grows of the condensate response, when the eigenmodes are excited. 
The temperature dependence of the cross-mobility drastically changes in this case.
We want to emphasize also, that even in dirty samples, where the diffusive regime of transport dominates, there might occur the exciton flux at low temperatures (with the obvious restriction that the exciton condensate is not destroyed by the impurity field).


\section*{Acknowledgements} 
This work was supported by the Foundation for the Advancement of Theoretical Physics and Mathematics BASIS Grant No.~17-15-526-1 (M.V.B), the Institute for Basic Science in Korea Project No.~IBS-R024-D1 (I.G.S). Authors acknowledge the support by the Russian Science Foundation Project No.~17-12-01039 of the analysis of the Coulomb drag in the case of Bose-condensed excitons.



\appendix

\section{Screening of electron-exciton interaction in normal and BEC phases of exciton gas}
\label{ApA}
Screening of external fields plays an important role in transport of particles in nanostructures~\cite{K_Ch_2}, in particular, in the Coulomb drag problem~\cite{Kamenev_Oreg}. 
Here we derive and analyze the formula for the dielectric permittivity of the electron-exciton system under study. 

Any fluctuation of particles density $\delta n$ induces an additional potential $\delta V$. For example, in the case of a one-component system, 
\begin{gather}\label{app.1a}
\delta V(\textbf{q},\omega)=\tilde{V}(\textbf{q})\delta n(\textbf{q},\omega),
\end{gather}
where $\tilde{V}(\textbf{q})$ is a bare inter-particle potential (further we will omit arguments of the Fourier transforms for simplicity).
Let us consider a two-component system with intra- and inter-subsystem interactions. 
In this case the induced potentials in each of the subsystems take the form
\begin{gather}\label{app.2a}
\delta V_1=\tilde{V}_{11}\delta n_1 +\tilde{V}_{12}\delta n_2,
\\\nonumber
\delta V_2=\tilde{V}_{21}\delta n_1 +\tilde{V}_{22}\delta n_2,
\end{gather}
where the subscripts distinguish the subsystems. Within the linear response model, the density fluctuations are proportional to the perturbation, thus
\begin{gather}\label{app.3a}
\delta n_1=\Pi_1(V_{11}+V_{12}),
\\\nonumber
\delta n_2=\Pi_2(V_{21}+V_{22}),
\end{gather}
where $\Pi_i$ is a response function and $V_{ij}=\tilde{V}_{ij}+\delta V_{ij}$ is a full interaction potential. 
Substituting Eq.~(\ref{app.3a}) in~(\ref{app.2a}) and taking into account that the induced potential indeed represents a sum of two terms $\delta V_i=\delta V_{ii}+\delta V_{ij}$, we derive the Dyson equation for the screened potentials,
\begin{gather}\label{app.4a}
\left(
        \begin{array}{cc}
           V_{11} & V_{12} \\
           V_{21} & V_{22} \\
        \end{array}
\right)=
\left(
        \begin{array}{cc}
           \tilde{V}_{11} & \tilde{V}_{12} \\
           \tilde{V}_{21} & \tilde{V}_{22} \\
        \end{array}
\right)+
\\\nonumber
+\left(
        \begin{array}{cc}
           \tilde{V}_{11} & \tilde{V}_{12} \\
           \tilde{V}_{21} & \tilde{V}_{22} \\
        \end{array}
\right)
\left(
        \begin{array}{cc}
           \Pi_1 & 0 \\
           0 & \Pi_2 \\
        \end{array}
\right)
\left(
        \begin{array}{cc}
           V_{11} & V_{12} \\
           V_{21} & V_{22} \\
        \end{array}
\right).
\end{gather}
From the solution of Eq.~\eqref{app.4a} it follows that the screened inter-subsystem interaction is proportional to the bare one $V_{12}=\tilde{V}_{12}/\epsilon$, where $\epsilon$ is the permittivity, which we are interested in.

At temperatures higher than the BEC transition temperature, the dielectric function reads 
\begin{gather}\label{app.0}
\epsilon^R(\textbf{q},\omega)= \left(1-v_q\Pi_{\textbf{q},\omega}^R\right) \left(1-gP_{\textbf{q},\omega}^R\right)
-V_q^2\Pi_{\textbf{q},\omega}^RP_{\textbf{q},\omega}^R,
\end{gather}
where $v_q=2\pi e^2/\epsilon q$ is the electron-electron interaction potential, $\Pi_{\textbf{q},\omega}$ and $P_{\textbf{q},\omega}$ are the electronic and excitonic polarization operators, respectively, and the interactions $g$ and $V_q$ have been introduced in the main text. 
We assume that the last term in~(\ref{app.0}) is negligible since $V_q^2/ v_qg=|q\rightarrow x/2l|=xe^{-x}d/4l <  d/l \ll 1$. Hence we will disregard it in what follows. Then the total dielectric function represents a product of the electronic and excitonic contributions:
\begin{gather}\label{app.1}
\epsilon^R(\textbf{q},\omega)=\epsilon_e^R(\textbf{q},\omega) \epsilon_{ex}^R(\textbf{q},\omega),
\end{gather}
where by definition
\begin{gather}\label{app.2}
\epsilon_e^R(\textbf{q},\omega)=1-v_q\Pi_{\textbf{q},\omega}^R,
\\\nonumber
\epsilon_{ex}^R(\textbf{q},\omega)=1-gP_{\textbf{q},\omega}^R.
\end{gather}
Explicit forms of these expressions depend on the regime of particles motion.
In the case of diffusive electron motion, the polarization operator reads~\cite{Kamenev_Oreg}
\begin{gather}\label{app.3}
\Pi_{\textbf{q},\omega}^R=-\frac{m}{\pi}\frac{Dq^2}{Dq^2-i\omega},
\end{gather}
where $D=v_F^2\tau_e/2$ is a diffusion constant of electrons, while in the ballistic case, we have 
\begin{equation}
\Pi_{\textbf{q},\omega}^R=-m/\pi.
\end{equation}

The permittivity of the exciton gas in normal phase is~\cite{K_Ch_1}
\begin{gather}\label{app.4}
\epsilon_{ex}^R(\textbf{q},\omega)\equiv\epsilon_T=1+4\frac{M}{m_e}\kappa d\left(e^{T_c/T}-1\right),\,\,
\end{gather}
where $T_c=\pi N/2M$.

To find the permittivity at low temperatures, it is necessary to be careful to avoid double counting of the inter-exciton interaction. 
Indeed, the propagators in Eq.~(\ref{eq.IV.4}) already  include it, so the permittivity reads
\begin{gather}\label{app.5}
\epsilon^R(\textbf{q},\omega)= \left(1-v_q\Pi_{\textbf{q},\omega}^R\right) 
-V_q^2\Pi_{\textbf{q},\omega}^R\mathcal{P}_{\textbf{q},\omega}^R,
\end{gather}
where
\begin{gather}\label{app.6}
\mathcal{P}_{\textbf{q},\omega}^R= \mathfrak{G}_{\textbf{q}}(\omega)^R +\mathfrak{F}_{\textbf{q}}(\omega)^R +\mathfrak{F}_{\textbf{q}}(\omega)^{+,R}
+\tilde{\mathfrak{G}}_{\textbf{q}}(\omega)^R
\\\nonumber
=\frac{g_sn_cq^2/M}{(\omega+i\gamma_{\textbf{q}})^2-\omega_{\textbf{q}}^2}
\end{gather}
is the linear response function of the excitonic BEC.
Note, that putting $b=1$ in order to perform integration in Eq.~\eqref{eq.IV.8} and~\eqref{eq.IV.13} we, in fact, disregard the second term in Eq.~\eqref{app.5}, like in the case of large temperatures. 
Indeed, with our parameters $b$ varies from $0.9$ to $1.0$, thus putting $b=1$ should not lead to a large error.

\

\end{document}